# Chapter 12

# Vacuum System

*V. Baglin\*, P. Chiggiato, P. Cruikshank, M. Gallilee, C. Garion and R. Kersevan*

CERN, Accelerator & Technology Sector, Geneva, Switzerland

## 12 Vacuum system

### 12.1 Overview

The luminosity upgrade programme (HL-LHC) requires modifications of the present LHC's vacuum system, in particular in the triplets and experimental areas. Such modifications must follow guidelines, similar to those followed for the present machine. The increased stored current implies a higher thermal power in the beam screen from the image current moving along with the stored particles and stronger synchrotron radiation (SR) and electron cloud (EC) effects, which in turn translate into higher degassing rates.

One of the main tasks of the vacuum HL-LHC work package is to define the geometry of the vacuum equipment in the new superconducting (SC) triplet and Separation dipole D1 magnets. It is also necessary to define a strategy for assembling and inserting high-density shielding material into the SC IR magnets. This is mandatory for protecting the magnets from collision debris coming from the experiments' interaction points (IPs). A balance between cold bore size and vacuum pumping system will be defined based on experience gained with the present machine and recent advances on new materials. A number of new ideas have recently emerged – for example, the amorphous carbon coating for which validation is ongoing.

The change of the aperture of the triplets at IR1 and IR5 implies that the experimental vacuum chambers of CMS and ATLAS will require a review of aperture, impedance, and vacuum (dynamic and static) values. From preliminary analyses, the forward regions of CMS and ATLAS will need to be adapted to cope with the new beam geometry in IR1 and IR5. New materials will likely be needed to mitigate the additional activation from the increased luminosity. New access procedures will be needed to allow the minimization of the integrated dose to personnel. With the HL-LHC, less flexibility will be available for the optics of LHCb and ALICE; therefore, the vacuum chambers at IR2 and IR8 must be validated for ultimate running conditions to ensure that these chambers do not impose a limitation. Positions of mechanical supports, pumps, and gauges must be analyzed to ensure that layouts are optimized for the new machine configuration. Bake-out equipment will be redefined depending on activation and specific needs. All experimental chambers must be treated with Non-Evaporative Getter (NEG) or equivalent to minimize secondary electron yield (SEY), thus reducing electron cloud effects.

### 12.2 Beam vacuum requirements

The HL-LHC beam vacuum system must be designed to ensure the required performance when beams with HL-LHC nominal parameters circulate. The system must be designed for HL-LHC ultimate parameters, without a margin.

The vacuum system must be designed to avoid pressure runaway induced by ion-stimulated desorption. It must also be designed to take into account the effects of synchrotron radiation, electron cloud, and ion-stimulated desorption from the walls. Heat load onto the beam vacuum chamber walls or flanges and beam impedance effects must also be taken into account [1].

---

\* Corresponding author: Vincent.Baglin@cern.ch

The system must be compatible with the global LHC impedance budget and the machine aperture.

The average gas density along the IR must satisfy the level defined by the 100 h vacuum lifetime due to nuclear scattering, i.e. less than $1.2 \times 10^{15}$ $H_2$ molecules m$^{-3}$ in the LHC [2]. This limit decreases proportionally to the inverse of the beam current. Table 12-1 gives the molecular gas densities yielding a 100 h vacuum lifetime in the LHC and the HL-LHC assuming the presence of a single gas in the vacuum system. The average gas density along IR1, IR2, IR5, and IR8 must also ensure that the background to the LHC experiments is at a minimum [3, 4]. In the absence of specified values from the LHC experiments themselves, the LHC design value will be scaled to HL-LHC parameters.

Table 12-1: Single gas species molecular gas density (m$^{-3}$) to satisfy 100 h vacuum lifetime in the LHC and the HL-LHC [2].

| Machine | I [A] | $H_2$ [m$^{-3}$] | $CH_4$ [m$^{-3}$] | $H_2O$ [m$^{-3}$] | CO [m$^{-3}$] | $CO_2$ [m$^{-3}$] |
|---|---|---|---|---|---|---|
| LHC | 0.58 | $1.2 \times 10^{15}$ | $1.8 \times 10^{14}$ | $1.8 \times 10^{14}$ | $1.2 \times 10^{14}$ | $7.9 \times 10^{13}$ |
| HL-LHC | 1.09 | $6.4 \times 10^{14}$ | $9.6 \times 10^{13}$ | $9.6 \times 10^{13}$ | $6.4 \times 10^{13}$ | $4.2 \times 10^{13}$ |

## 12.3 Vacuum layout requirements

The vacuum layout must ensure the vacuum requirements when beams with HL-LHC nominal parameters circulate. The system must be designed for the HL-LHC ultimate luminosity (i.e. $7.5 \times 10^{34}$ cm$^{-2}$ s$^{-1}$), without margin.

- All beam vacuum elements must be leak tight (leak rate less than $10^{-11}$ mbar$^{-1}$/s He equivalent) clean according to CERN vacuum standards and free of contamination.

- According to the LHC baseline [2], the vacuum system in the LSS must be sectorized with gated valves. The vacuum sectorization is delimited by cold-to-warm transitions, length of vacuum sectors, or specificity of components (fragility, maintenance, etc.) [5].

- Vacuum sector valves must be installed at each cold-to-warm transition in order to decouple the room temperature and cryogenic temperature vacuum systems during bake-out and cool-down phases.

- The distance between the vacuum sector valve and the cold-to-warm transition must be minimized in order to reduce the length of the beamline that is not baked-out in situ.

- Dedicated vacuum instrumentation must be provided close to and either side of each sector valve and along each vacuum sector.

- Sector valves must be remotely controlled and interlocked in order to dump the circulating beam in the case of malfunctioning. The LHC and HL-LHC vacuum sectorizations delimit two types of vacuum system:
    o room temperature vacuum system;
    o cryogenic temperature vacuum system.

- The vacuum system shall be integrated in the tunnel and cavern volumes with the permanent/mobile bake-out system, bake-out racks, quick flanges collars, mobile pumping systems, and diagnostics systems. The corresponding space must be reserved into the tunnel integration to allow a proper access and operation of the vacuum system.

- Integration studies must also be performed for installation and un-installation phases of equipment to identify potential conflicts.

- Integration and installation drawings must be circulated and validated before installation in the tunnel and caverns.

- The vacuum chamber aperture is defined by the beam optics system, and by machine protection and background for experimental considerations. The aperture of the vacuum chamber must not be the limiting aperture.

- All components to be installed into the vacuum systems must be approved and their vacuum performance validated before installation.

- A maximum number of LHC beam vacuum components will be reused for the HL-LHC upgrade.

- High radiation areas along the LSS must be clearly identified at an early stage of the design, in particular to highlight positions where remote handling/tooling might be preferred and positions where instrumentation must be radiation resistant.

- When needed, irradiation tests of specific components (instruments, bake-out jackets, cables, electronics, O rings, etc.) must be conducted to meet the radiation dose specifications.

- The spares policy will follow the general A&T sector policy. A spares policy must be made available from early on during procurement to benefit from large quantity orders, in particular for highly specialized equipment such as beam screens, modules, etc.

- Cryogenic elements must be installed first; then room temperature vacuum sector valves, followed by completion of the room temperature vacuum sectors.

- Time, resources, and space to allow the temporary storage of LHC vacuum components, which need to be dismounted to allow HL-LHC infrastructure modifications and equipment installation, will be evaluated in the next phase of the project.

### 12.3.1 Room temperature vacuum system requirements

Standard vacuum chambers and vacuum modules connect the machine components.

In order to accommodate thermal expansion during bake-out and sustain 'vacuum forces' due to the differential pressure, each component containing a beam pipe must have a single fixed point. Other supporting points, if any, must be sliding.

All machine components operating at room temperature must be bakeable. The required bake-out temperature is 230°C ±20°C for NEG coated vacuum chambers and 300°C ±20°C for uncoated stainless steel beam pipes. To optimize the bake-out duration while minimizing the mechanical strength and radiation to personnel, the applied heating rate during bake-out is 50°C/h.

The vacuum system layout must be designed to fulfil the stated requirements throughout a full run. In particular, the impact of the outgassing rate of specific components, e.g. collimators, must be taken into account during the layout definition phase. No NEG coating re-activation can be foreseen during short stops to recover loss of pumping performance.

A vacuum module equipped with a bellows must be inserted between each machine component and vacuum chamber, and between the vacuum chambers themselves.

These vacuum modules must be screened by an RF bridge for impedance reasons.

For the sake of cost, reliability, spares policy, and standardization, the maximum number of vacuum module variants must be reduced with respect to the LHC baseline.

The warm vacuum chambers must be circular and bakeable. The current LHC variants are 80 mm, 100 mm, 130 mm, and 212 mm ID: any further variants needed for the HL-LHC will be kept to the minimum necessary.

Special chambers may be designed if needed but the quantity must be kept to a minimum. A typical case is that of the recombination chambers installed into the TAN absorber, which by definition is not circular.

Vacuum chamber transitions (VCT), which gives offsets and adaptation between pipe apertures, must be integrated at the early design stage into the concerned equipment by the equipment owners themselves, e.g. BTV, ACS, TCDQ, etc. in agreement with the vacuum group.

The vacuum chambers are aligned by TE-VSC within ±3 mm accuracy. Better tolerance will require the installation of survey targets. Alignment of other equipment is the responsibility of the survey group.

The choice of the vacuum chamber material between Cu alloy and stainless steel (either Cu coated or bare) is dictated by beam impedance constraints. Aluminium alloys are preferred in high-radiation areas

Connections between equipment must be made by Conflat bolt technology unless radiation issues and/or remote handling require the use of quick-release flanges with, for example, chain clamps.

The vacuum system must be integrated in the tunnel with the permanent or mobile bake-out system, bake-out racks, quick-flange collars, mobile pumping systems, and diagnostics systems. The corresponding space must be reserved during the tunnel integration studies to allow proper operation of the vacuum system.

### 12.3.2 Cryogenic temperature beam vacuum system requirements

The cryogenic beam vacuum system must be tightly decoupled by sector valves from the room temperature vacuum system. Dedicated instruments must be provided close to the sector valves to allow roughing, monitoring, and safety of the vessel.

A cold-to-warm transition must be integrated into the cryogenic beam vacuum sector at each extremity of the cryogenic system.

A continuous cold bore, i.e. without penetrating welds between the beam vacuum and helium enclosure, ensures leak-tightness between the superfluid helium and beam vacuum along the cryogenic beam vacuum sector. The LHC nominal cold bore temperature is 1.9 K in the arcs.

A beam screen must be inserted into the cold bore to extract the beam-induced heat load at a temperature higher than 1.9 K. The beam screen must be perforated with slots (4% transparency) to allow pumping into the cold bore space [2]. The LHC beam screen operates from 5 K to 20 K. The HL-LHC beam screens of the IT + D1 will probably need to run at a higher temperature (between 40 K and 60 K) due to a much higher heat load. In situ heating of the beam screen up to 90 K, with cold bore <3 K, is required to flush the condensed gas present on the beam screen inner surfaces towards the cold bore. This heating cycle may be required after a long technical stop or even between physics fills. The HL-LHC beam screen perforation percentage will be scaled to HL-LHC parameters and therefore increased as compared to the LHC.

When a cold bore operates at 4.5 K, cryoabsorbers are installed outside the beam screen in order to provide hydrogen pumping speed and capacity.

In the LHC, the maximum length without beam screen is less than 1 m. This LHC design value will be scaled to HL-LHC parameters and therefore reduced.

For the HL-LHC, the beam screen aperture will be derived from beam optics and magnet aperture inputs.

### 12.4 Insulation vacuum requirements

The insulation vacuum system ensures the required performance of the cryogenic system by eliminating the heat losses due to gas convection. The insulation vacuum systems under the responsibility of TE-VSC include the cryogenic distribution line (QRL) and cryogenic machine components, but exclude transfer lines outside the LHC tunnel and those of the experimental cavern.

The requirements of the insulation vacuum system for the HL-LHC can be summarized as follows:

- the pressure must be below $10^{-5}$ mbar;
- the helium leak rate, at the component level, must be below $10^{-10}$ mbar L/s;

- it must be compatible with the LHC insulation vacuum system [2];
- it must be built with the same standards used for the LHC insulation vacuum system.

The QRL and the magnet cryostats are connected via jumpers. However, the insulation vacuum of the QRL and continuous cryostat is sectorized through vacuum barriers. There is no sectorization of the QRL in the LSS of the LHC. Sectorization of the HL-LHC cryostats shall ensure that longitudinal leak location techniques can be employed. Connection to cryo-plant or transfer lines outside the LHC tunnel shall be delimited by vacuum barriers.

The insulation vacuum relies on cryopumping in normal operation. Fixed turbomolecular pumping groups are used for pumping before cool-down. This system also mitigates the impact of helium leaks during operation. Such pumps are also used for the detection of helium or air leaks. Dedicated pumping ports are used for rough pumping groups, pressure gauges, pressure relief valves, longitudinal leak localization techniques, and additional pump placement in case of operational leaks. A bypass equipped with isolation valves is installed between subsectors. The standard for pumping ports is the ISO-K DN 100 flange. Each insulation vacuum volume has to be equipped with pressure relief valves. Elastomer seals (Viton, NBR) are used where system demountability is necessary (interconnections, instrumentations, etc.). The layout detailed design, testing, and final acceptance of the HL-LHC cryostats are subject to TE-VSC approval.

For the HL-LHC project, in the areas of expected high irradiation, specific seals (metals or hard-rad polymers) have to be installed on new equipment and be used to replace standard seals on any retained equipment.

Regular preventive maintenance will be carried out on turbomolecular pumping groups during technical stops.

## 12.5 Experimental vacuum system requirements

The experimental vacuum system is located between Q1L and Q1R of each interaction point. Similarly to the LSS, the vacuum layout of each experimental vacuum system must ensure the vacuum requirements when beams with HL-LHC nominal parameters circulate. The system shall be designed for HL-LHC ultimate parameters, without margin. Therefore, all constraints and requirements defined in Sections 12.3 and 12.3.1 apply in this section.

During long beam stops (>10 days), neon venting is needed to protect the fragile experimental vacuum chambers from deformations caused by mechanical shocks. The baseline is that there will be no work in the vicinity of the vacuum chambers while they are under vacuum.

The vacuum chamber supporting system must be compatible with standard activities performed in the experimental cavern during short stops (e.g. winter technical stops). In particular, no personnel are allowed in the vicinity of the beam pipe (<2 m radius) during these phases.

A vacuum sector valve is installed at each Q1 extremity. This vacuum sector valve, installed just after the cold-to-warm transition, ensures decoupling of the room temperature and cryogenic temperature vacuum systems during bake-out and cryogenic temperature transients. The Q1 sector valves are interlocked to the circulating beam.

A second vacuum sector valve is installed between Q1 and the TAXS on the lefthand and righthand sides of the ATLAS and CMS experiments. On the lefthand and righthand sides of the ALICE and LHCb experiments a sector valve is installed between Q1 and the cavern shielding. Such vacuum valves allow the decoupling of two delicate and complex beam vacuum sectors, i.e. the inner triplets' vacuum sector and the experimental vacuum sector, in areas that can be radioactive. They are blocked open during operation, i.e. they are out of the interlock chain.

A rupture disk is installed in the buffer zones in order to protect the experimental vacuum chambers in case of liquid helium inrush. Therefore, the Q1 sector valve located at the cold-to-warm transition position must be interlocked against possible helium inrush in case of a cryogenic accident inside the inner triplets.

As for the present LSS vacuum system, all machine components operating at room temperature must be bakeable and NEG coated.

Scheduled or accidental air venting in air for repair or maintenance of any of the vacuum sectors of the experimental vacuum system implies a complete NEG recommissionning of the beam pipes, i.e. two bake-out cycles, the first for the bake-out of the metallic part, the second for NEG activation.

An ultra-pure neon venting system is installed in the buffer zone (for ATLAS, CMS, and ALICE) or in the detector itself (i.e. the vertex locator, VELO) to allow remote venting of the experimental vacuum system during long stops (>10 days).

### 12.5.1 High luminosity experiments: ATLAS and CMS

ATLAS and CMS vacuum layout drawings are described in Refs. [6, 7].

On both sides of ATLAS and CMS, a pumping system is installed in the buffer zone to allow pump-down and vacuum commissioning during NEG activation of the ATLAS and CMS experiment.

In the Q1–TAXS areas instrumentation must be minimized. The Q1 sector valve must remain in the interlock chain in order to protect the experimental area from gas contamination or liquid helium inrush.

The following are required to avoid personnel intervention in a high radiation area.

- A pumping and neon systems must be installed in the buffer zone on the righthand side of ATLAS to complement the lefthand side.
- Remote tooling must be foreseen to avoid personnel intervention. Quick type flanges are mandatory. Welds are preferred to flanges.
- Installed components must be robust: in particular, sliding fingers in RF bridges are forbidden.
- The bake-out system must be permanent and fully integrated with the other systems from the design phase.
- Detector modifications shall be designed taking into account movement and intervention of personnel. They shall also take into account the beam pipe system and flanges for detector movements. It must be ensured that standard interventions during technical stops and chamber replacements are thoroughly studied during the design phase with the aim of minimizing the radiation dose to personnel during the detector and chamber lifecycle.

ATLAS is equipped with a permanent bake-out system while CMS is not. However, the HL-LHC might require the use of permanent bake-out systems for both detectors due to radiation issues. In CMS, a specific design of the jackets, thermocouple, and cable layouts must also comply with radiation requirements.

The vacuum modules are not screened by sliding RF finger for access and space constraints, radiation protection reasons, and potential risk of aperture loss.

The alignment of the vacuum chambers shall be performed remotely with appropriate hardware.

The choice of the vacuum chamber material between Cu alloy and stainless steel (either Cu coated or bare) is dictated by beam impedance requirements. Al alloys are preferred in high-radiation areas.

The connection between equipment shall be done by Conflat bolted technology unless there are radiation issues and/or remote handling that requires the use of quick-release flanges.

It is assumed that the ATLAS central beam pipe inner diameter, as installed during LS1, remains the same until at least LS3 [8].

According to Ref. [8], that the following are assumed.

- The CMS central beam pipe inner diameter, as installed during LS1 remains the same until at least LS3.
- End cap, HF, CT2, and forward pipes of the CMS vacuum system must be upgraded to Al bulk material during LS2. No mechanical intervention between TAXS L and TAXS R is therefore expected during LS3.

Any new chamber to be installed in the future must be compatible with remote tooling, e.g. equipped with a 'quick flange' type system.

Since the TAS needs to be replaced during LS3, the vacuum system located inside the experimental cavern needs to be recommissioned, i.e. NEG activation even if no changes to the vacuum system inside the cavern are foreseen.

### 12.5.2 ALICE and LHCb experiments

On the righthand side of ALICE, a pumping system is installed in the buffer zone to allow pumpdown and commissioning during NEG activation of the beam pipes.

In the ALICE cavern, a manual valve, located on the lefthand side of the beryllium central beam pipe, allows the isolation of the detector vacuum sector from the RB24 vacuum sector. Consequently, the vacuum chambers in the RB24 area can be dismounted during long shutdowns if needed.

The ALICE vacuum layout drawing is described in Ref. [9].

The ALICE central beryllium beam pipe is not equipped with a permanent bake-out system. The beam pipes located in the RB26 area of the ALICE cavern (i.e. the righthand side of the IP) are equipped with a permanent bake-out system. From the left of the central beam pipe to the end of RB26 the vacuum modules are not screened by sliding RF fingers because of access and space constraints, radiation issues, and the potential risk of aperture loss.

In LHCb, the vertex locator (VELO) detectors are installed into a secondary vacuum system, which is isolated from the beam vacuum system by a thin RF shield. The RF shield mechanical integrity is protected during pumping and venting phase by an automatic gas balance system to maintain the pressure difference within a mechanically stable range.

At the VELO, an automatic pumping system is installed to allow pumpdown and commissioning during NEG activation while protecting the VELO RF shield.

At the VELO, an ultra-pure neon venting system is installed to allow remote venting of the LHCb with an automatic gas balance system while protecting the VELO RF shield during long stops (>10 days).

The LHCb beam pipes are not equipped with a permanent bake-out system. However, it is foreseen that part of the UX85/3 vacuum chamber in the region of the calorimeter will be equipped with permanent bake-out for LS2 [10].

The VELO beam pipe must be baked for 48 h at a minimum of 180°C to suppress the electron cloud.

From the righthand side of the VELO beam pipe to the end of UX85, the vacuum modules are not screened by sliding RF fingers because of access and space constraints, radiation issues, and the potential risk of aperture loss.

Wake-field suppressors are installed at the VELO beam pipe extremities for impedance reasons.

The LHCb vacuum layout drawing is described in Ref. [11].

According to Ref. [12], it is assumed that the ALICE central beam pipe inner diameter will be changed during LS2 and remains unchanged afterward.

According to Ref. [10], is the following are assumed that.

- The LHCb VELO beam pipe will be changed during LS2 and remains in the ring afterward [3].
- The 2 m long Cu-alloy vacuum chamber, VCDBV, located on the righthand side of the VELO will be replaced by an Al alloy beam pipe. This change will take place during LS2. No modification is expected afterward.
- Part of the UX85/3 vacuum chamber in the calorimeter region will be equipped with permanent bake-out jackets during LS2. It is also assumed that there will be no more installation of permanent bake-out jackets after LS2.

**12.6 Beam screen requirements**

Beam screens are inserted into cryogenic cold bores in order to guarantee vacuum performance. They are part of the LHC vacuum system baseline [1]. The requirements of the HL-LHC beam screen can be summarized as follows.

- The system lifetime must be longer than the HL-LHC lifetime, i.e. 40 years of operation and 50 quenches.
- The system must fulfil the required vacuum performance in terms of vacuum stability, vacuum lifetime, and background to the experiments.
- The shielded beam screen must be perforated (or be shadowing) in order to fulfil the vacuum performance in a way to allow the pumping of molecules onto the cold bore.
- The cold bore temperature must be held below 3 K to allow cryosorption of all molecules with the exception of helium.
- The shielded beam screen must be heatable to allow a transfer of the gas onto the shielded beam screen towards the cold bore during machine stops.
- The shielded beam screen must withstand the Lorentz forces induced by eddy currents during a quench.
- The temperature of the shielded beam screen must be actively controlled in a given temperature range.
- To minimize the heat load on the shielded beam screen due to the electron cloud, a coating, e.g. amorphous carbon or a clearing electrode system, must be studied, validated, and implemented. The e-cloud mitigation measures must be applied not only to the new high-luminosity insertion regions, IP1 and IP5, that will renovated. They will be applied also to the interaction regions of IP2 and IP8. These are indeed low-luminosity insertions, however the beam pattern will be the same as in the high-luminosity insertions. This point is quite recent and mitigation measures are currently under investigation.
- The system must be compatible with the impedance budget.
- The system must be compatible with the machine aperture.
- The surface of the beam screen must withstand a dose of several hundred MGy during its lifetime.

In order to operate properly, the beam vacuum system must be evacuated for five consecutive weeks, at least, prior to cool-down to allow the outgassing rate of adsorbed water to be reduced to acceptable levels.

During cool-down of a cryogenic system, the cold bore must be cooled first in order to minimize gas condensation onto the beam screen.

In the case where gas condenses onto the beam screen during operation, e.g. after a magnet quench, a transfer of this gas towards the cold bore via beam screen heating is needed. This procedure should be carried out in a couple of days.

The HL-LHC beam screen must be inserted during the cryostating phase prior to tunnel installation. The surface of the beam screens must be kept clean during assembly. This implies that the beam screen is installed at the last stage of cryostating. Without specific tooling and procedures, no probe or device can be inserted into the vacuum system once the beam screens are installed.

The cooling tubes must be dimensioned to allow a proper cooling of the system during operation within the limits defined above.

According to vacuum standards, full penetrating welds are forbidden in the vessel wall separating the beam vacuum and helium enclosures. However, given the limited number and length of new vacuum elements to be installed, this general rule might be revised in the case of compelling reasons, if compensated for by adequate measures of risk mitigation.

Depending on the location, two types of beam screens may exist: shielded and non-shielded beam screens. The shielded beam screens intercept part of the debris produced at the high luminosity IPs, thereby protecting the cold masses from radiation-induced damaged.

Amorphous carbon (a-C) coating is at present the baseline for the inner surface of the HL-LHC shielded and non-shielded beam screens. Due to a-C's properties, strong electron cloud suppression in these HL-LHC components is expected. Amorphous carbon coating will be applied to HL-LHC beam screens if needed for the reduction of heat load to cryogenic systems, reduction of background to experiment, and/or beam physics requirements.

A demonstration of the electron cloud suppression must be performed with a dedicated set up such as COLDEX in SPS LSS4. Studies will be held during SPS scrubbing runs and dedicated machine developments. Experimental results must be supported by theoretical expectations derived from simulations codes such as Ecloud or PyCloud.

Amorphous carbon coating should be the last step of beam vacuum preparation before lowering the magnet into the tunnel, avoiding any subsequent insertion of tooling or other devices into the beam vacuum line.

For IP2 and IP8, in situ coating of the present beam screen (or alternatively the placement of a clear electrode) must be studied and conducted during the long shutdown (LS2 or LS3). If in situ coating is not possible, removal of the magnet cryostat will be considered to allow beam screen exchange.

If needed, a sawtooth structure can be produced on the beam screen walls (dipoles and quadrupoles). The sawtooth must be designed to intercept the synchrotron radiation at perpendicular incidence and to reduce the forward scattering of light.

12.6.1    Shielded beam screen

HL-LHC shielded beam screens are to be inserted into HL-LHC D1 and IT of LSS1 and LSS5. These beam screens ensure the vacuum requirements, the shielding of the cold mass from physics debris, and the screening of the 1.9 K cold bore cryogenic system from beam-induced heating.

As a baseline, the shielded beam screen is assumed to fulfil the vacuum requirements with a-C coating operating at 40–60 K.

The shielded beam screen is estimated to operate in the 40–60 K ±10 K temperature range. The operating temperature range will be defined by the available cooling power, the expected beam-induced heat load, and compatibility with the vacuum requirements.

The selected shielding material is Inermet, a tungsten alloy. It is made of 40 cm long blocks, which must be accommodated on the beam screen structure [13].

12.6.2  Non-shielded beam screen

HL-LHC non-shielded beam screens are to be inserted into HL-LHC D2, Q4, Q5, and Q6 of LSS1 and LSS5 and, if needed, in D1, DFBX, and IT of LSS2 and LSS8. Such beam screens ensure the vacuum requirements together with screening of the 1.9 K cold bore from beam-induced heating.

As a baseline, the a-C beam screen is assumed to fulfil its vacuum requirements with a-C coating operating at 5–20 K.

In LSS1 and LSS5, the a-C beam screen is part of the new D2, Q4, Q5, and Q6 system for the HL-LHC, which will replace the present ones.

In LSS2 and LSS8, the a-C coated beam screens will replace part of the present beam screens. The dismounting of the D1, DFBX, and IT of LSS2 and LSS8 is needed unless the cryogenic system can be upgraded.

The a-C coated beam screen is estimated to operate in the 5–20 K ±3 K temperature range. The operating temperature range will be defined by the available cooling power, the expected beam-induced heat load and compatibility with the vacuum requirements.

The selected a-C coated beam screen material is P506 non-magnetic stainless steel. It is similar to that used for the present beam screens. Copper will be co-laminated for impedance reasons. The proposed thickness of the Cu layer is 60 μm.

12.6.3  Vacuum beam line interconnection

HL-LHC shielded beam screens are to be inserted into HL-LHC D1 and IT of LSS1 and LSS5. Beam vacuum interconnections ensure the continuity of the beam vacuum envelope, a smooth transition between adjacent beam screens, and the electrical continuity of the image current. The beam screens are fixed on one side to the cold mass; on the other side, a compensation bellows between the beam screen extremity and the cold mass has to be integrated to cope with the differential thermal displacements between the beam screens and the cold mass.

The HL-LHC beam screens rely on cooling tubes larger than those currently used. The routing of these cooling tubes, in and out of the cold bore, requires exit pieces with significant longitudinal space. Trough-wall welds on the helium circuit are forbidden in the beam vacuum. Automatic welds have to be used in the insulation vacuum.

A bellows between two adjacent beam screens has to be integrated to allow thermal contraction as well as to compensate for the mechanical and alignment tolerances. In the transitions, collisions debris shielding with a circular aperture and non-sliding RF fingers are foreseen and are being studied.

The vacuum beamline interconnections in the triplets integrate BPM as well. They define the interconnection length. Therefore, iteration and optimization of the different elements are required to complete machine integration.